# Melting and decomposition of orthorhombic $B_6Si$ under high pressure

Vladimir L. Solozhenko,[1,*] Vladimir A. Mukhanov [1] and Vadim V. Brazhkin [2]

[1] *LSPM–CNRS, Université Paris Nord, 93430 Villetaneuse, France*

[2] *Institute for High Pressure Physics, Russian Academy of Sciences, 108840 Troitsk, Russia*

**Abstract**

Melting of orthorhombic boron silicide $B_6Si$ has been studied at pressures up to 8 GPa using *in situ* electrical resistivity measurements and quenching. It has been found that in the 2.6–7.7 GPa range $B_6Si$ melts congruently, and the melting curve exhibits negative slope of -31(2) K/GPa that points to a higher density of the melt as compared to the solid phase. At very high temperatures $B_6Si$ melt appears to be unstable and undergoes disproportionation into silicon and boron-rich silicides $B_nSi$ ($n \geq 12$). The onset temperature of disproportionation strongly depends on pressure, and the corresponding low-temperature boundary exhibits negative slope of -92(3) K/GPa which is indicative of significant volume decrease in the course of $B_6Si$ melt decomposition.

## 1. Introduction

The boron–silicon system − despite of rather long research history − remains not fully understood even at ambient pressure [1]. There are three main groups of B–Si binary compounds usually referred to as "$B_3Si$" (with the stoichiometry ranging from $B_{2.8}Si$ to $B_{4.8}Si$); $B_6Si$; and boron-rich silicides i.e. $B_nSi$ ($12 \leq n \leq 50$) (for details see [2] and references therein). Boron silicides attract considerable attention due to superior thermal stability, excellent chemical resistance, promising mechanical and electronic properties that offers potential for their use as advanced engineering and smart functional materials. The boron-silicon phase diagram at ambient pressure is rather complicated and includes a number of peritectic reactions and intermediate phases of different stoichiometries [3]. As for the phase relations in the B–Si system at high pressures and high temperatures, they have not been studied at all.

$B_6Si$ is the most explored boron-rich silicide that is used in nuclear technology [4], some nanotechnology applications [5] and as high-temperature thermoelectric material [6]. It has an orthorhombic unit cell (space group *Pnnm*) containing 43 silicon atoms and 238 boron atoms forming a dense framework of 18 icosahedra and 4 icosihexahedra [7] (Fig. 1*a*). According to the predictions made in the framework of thermodynamic model of hardness [8], $B_6Si$ is expected to exhibit hardness of about 35 GPa [9] comparable to that of polycrystalline boron carbide. High pressure – high temperature behavior of $B_6Si$ has not been studied so far. In the present work we have performed the first investigation of solid and liquid $B_6Si$ at pressures up to 8 GPa.

---

[*] vladimir.solozhenko@univ-paris13.fr

## 2. Experimental

Powder (particles size less than 45 μm) of orthorhombic $B_6Si$ was used as received from ABCR GmbH (Germany). High-pressure experiments in the 2–8 GPa range have been performed using a toroid-type high-pressure apparatus with a specially designed high-temperature (up to 3500 K) cell [10,11]. The cell was pressure-calibrated at room temperature using phase transitions in Bi (2.55 and 7.7 GPa), PbSe (4.2 GPa), and PbTe (5.2 GPa). The temperature calibration under pressure was made using well-established reference points: melting of Si, NaCl, CsCl, Pt, Rh, $Al_2O_3$, Mo and Ni-Mn-C ternary eutectic. $B_6Si$ powder was compacted into pellets and placed in boron nitride (Saint-Gobain, grade AX05) capsules. After isothermal holding time of 60-300 s at desired pressure and temperature the samples were either quenched by switching off the power or stepwise cooled down to room temperature, and then slowly decompressed down to ambient pressure. In some experiments no BN capsule has been used (the sample was in a direct contact with graphite heater), and the appearance of a liquid phase upon heating could be detected *in situ* by electrical resistance measurements using the method described earlier [12-14]. In a special set of experiments it was found that $B_6Si$ does not react with BN and graphite in the whole studied pressure–temperature range.

The samples recovered from high-pressure experiments have been studied by powder X-ray diffraction (Equinox 1000 Inel diffractometer; Cu Kα and Co Kα radiation). The characteristic diffraction patterns are presented in Fig. 2.

## 3. Results and Discussion

Melting of orthorhombic boron silicide $B_6Si$ has been studied in the 2.6-7.7 GPa pressure range using *in situ* electrical resistivity measurements and quenching (the results are shown in Fig. 3). The melting curve exhibits negative slope of -31(2) K/GPa that points to a higher density of the melt as compared to the solid phase.

The electrical conductivity of $B_6Si$ sample before melting is comparable to that of graphite heater at the same temperature, which corresponds to the literature data [15], taking into account a possible decrease in semiconductor band gap and an conductivity increase at high pressure. Upon melting, a slight (several percent) increase in sample resistance was observed. Melting occurs in a fairly narrow temperature range, comparable to the T-gradient value in the sample. In the samples quenched from the melt at temperatures slightly above the melting curve, no traces of any other phases were found except for orthorhombic $B_6Si$, i.e. the melting is congruent. It should be noted that according to the literature, at ambient pressure $B_6Si$ melts incongruently, and the width of the melting interval differs from one publication to another [3,16,17]. Extrapolation of the melting line obtained by us to the low-pressure region gives the value of 2110(20) K for melting point at ambient pressure, which is in good agreement with $B_6Si$ peritectic point (2123 K) according to the assessed equilibrium phase diagram of the B–Si system [3].

Quenching experiments with significant overheating of the melt relative to the melting curve resulted in decomposition of $B_6Si$ into silicon and boron-rich silicides. X-ray diffraction patterns of decomposition products are rather complex (see Fig. 2), but it is obvious that formation of neither $B_6Si$ and $B_3Si$, nor pure boron was observed. The main diffraction lines (in addition to reflections of crystalline silicon) can be attributed to $B_{14}Si$, $B_{36}Si$ (Fig. 1b), and $B_{50}Si$ compounds and/or their mixtures (note that diffraction patterns of these compounds differ significantly from one publication to another). The onset of $B_6Si$ decomposition at different pressures is observed at different degrees of overheating relative to the melting point: from 450 K at 2.7 GPa to 150 K at 7.7 GPa (Fig. 3). The melt decomposition boundary has a large negative slope of –92(3) K/GPa, which indicates that the decomposition is accompanied by a significant volume decrease. When extrapolating the lines of melting and decomposition to the high-pressure region, their intersection is observed at ~10 GPa, which suggests that at higher pressures $B_6Si$ decomposition into silicon and boron-rich silicides should be observed already in solid state.

The instability of $B_6Si$ under pressure with respect to disproportionation into silicon and boron-rich silicides can be explained by analyzing the volume effects of two hypothetical reactions

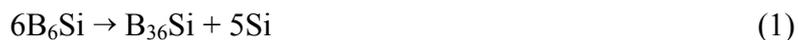
$$6B_6Si \rightarrow B_{36}Si + 5Si \qquad (1)$$

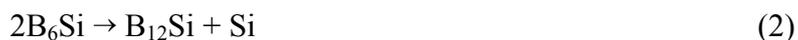
$$2B_6Si \rightarrow B_{12}Si + Si \qquad (2)$$

For convenience and simplification of the assessment of the second reaction, the $B_{12}Si$ stoichiometry rather than $B_{14}Si$ was taken which, however, does not affect the final result. Under normal conditions, the densities of the phases involved are as follows: 2.33 g·cm$^{-3}$ for Si; 2.42-2.44 g·cm$^{-3}$ for $B_6Si$ [7]; 2.48-2.51 g·cm$^{-3}$ for $B_{14}Si$ [18]; and 2.34-2.37 g·cm$^{-3}$ for $B_{36}Si$ [19]. It is easy to see that under normal conditions reaction (1) proceeds with a significant increase in volume, while in reaction (2) the volume practically does not change, i.e. both reactions cannot be stimulated by pressure.

The situation, however, changes drastically at high pressures and high temperatures. In the temperature range under study thermal expansion leads to an increase in the specific volume (by 3-4% [20]) for all compounds formed as a result of the considered disproportionation reactions. On the other hand, melting of silicon leads to its density increase by 10%, while the melting of boron-rich solids usually occurs with the preservation of $B_{12}$ icosahedra as the main structural units, and thus the volume change does not exceed 1-3%. In addition, all considered boron-rich silicides have bulk moduli of 170-190 GPa [21], which is twice as high as that of liquid silicon[†]. As a result, at high temperatures reaction (1) occurs with 1-3% volume decrease at pressures above 2 GPa; and reaction (2) occurs with a significant (5-7%) volume decrease in the whole pressure range under study. Thus, the observed large negative slope of the $B_6Si$ decomposition line is in full agreement with the thermodynamic estimations.

At ambient pressure $B_{14}Si$ compound is a semiconductor with a high resistivity, but at high temperatures (~2000 K) it has a semi-metallic conductivity of the order of $10^2$ $Ohm^{-1} \cdot cm^{-1}$ [24]. There is no data on electrical conductivity of rhombohedral $B_{36}Si$, however, it can be assumed that this compound is close in properties to β-rhombohedral boron. The decomposition of $B_6Si$ in our experiments, as a rule, was accompanied by an increase in conductivity, which is obviously associated with the very high conductivity of liquid silicon.

Note that the assumption about the possibility of solid-state decomposition of $B_6Si$ at pressures above 10 GPa does not contradict our estimates, since silicon at high (above 11 GPa) pressures passes into the high-pressure phase with a white-tin structure (SiII). This transformation in crystalline silicon is accompanied by ~20% volume decrease [23], therefore, the thermodynamic stimulus of reactions (1) and (2) at high pressures can only increase.

## 4. Conclusions

Summarizing, we can conclude that in the 2.6–7.7 GPa pressure range $B_6Si$ melts congruently, and both the solid phase and the melt have a bad-metal conductivity. Melting is accompanied by a small drop in electrical conductivity and a slight volume decrease. At significant overheating relative to the melting line, the decomposition of $B_6Si$ into silicon and boron-rich silicides is observed, and the onset temperature of decomposition decreases strongly with increasing pressure. In the future, it would be of interest to study $B_6Si$ up to the megabar pressure range. Apparently, at ultrahigh pressures and room temperature the diffusion processes are strongly suppressed, and decomposition of $B_6Si$ can occur at nanoscale only resulting in solid-state amorphization.

**Acknowledgements**

The authors are grateful to Dr. P.S. Sokolov for his assistance at the initial stage of work. The research was financially supported by the European Union's Horizon 2020 Research and Innovation Programme under Flintstone2020 project (grant agreement No 689279).

---

[†] In fact, bulk modulus of liquid Si has never been measured. 300-K bulk modulus of crystalline silicon makes 98 GPa, while in vicinity of melting point it drops down to 80-85 GPa [22]. Since Si melting curve under pressure is linear [23], one can conclude that at melting temperature bulk moduli of crystalline and liquid silicon should be very close.

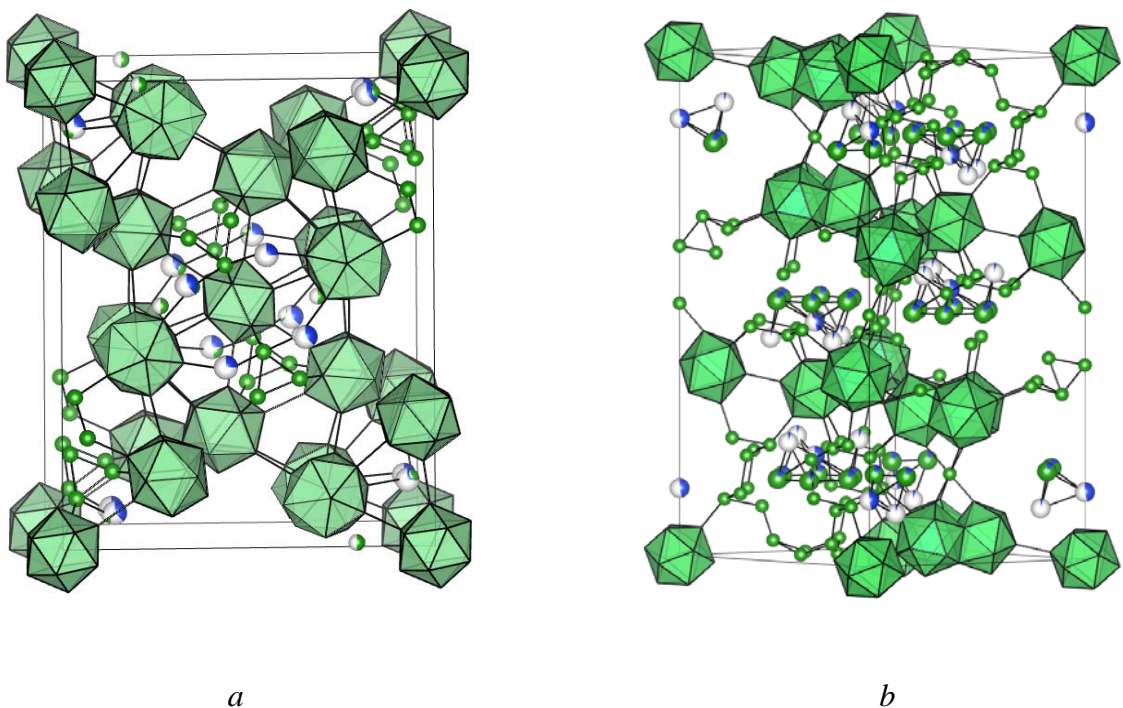

        *a*          *b*

Fig. 1 Crystal structures of orthorhombic $B_6Si$ [6] (*a*) and rhombohedral $B_{36}Si$ [19] (*b*). $B_{12}$-units are presented by green icosahedral polyhedral; silicon and boron atoms are presented by blue and green balls, respectively.

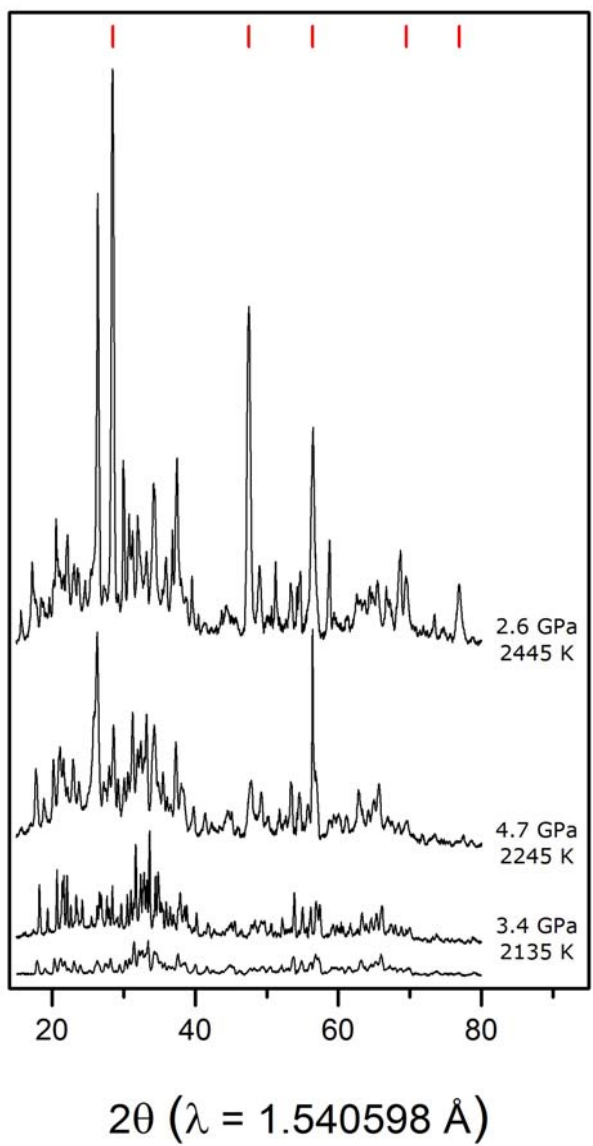

Fig. 2 X-ray diffraction patterns of pristine orthorhombic $B_6Si$ (bottom) and samples quenched from melt at different pressures and temperatures. Vertical red ticks correspond to positions of diffraction lines the cubic (*Fd-3m*) silicon.

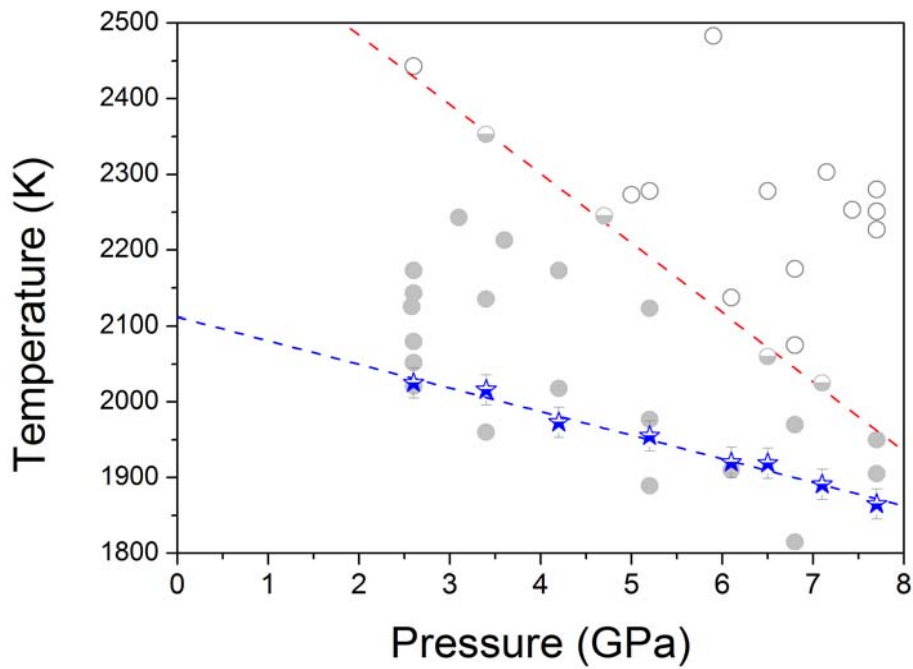

Fig. 3  $B_6Si$ melting and decomposition under pressure. Half-filled stars show the melting onset registered *in situ* by electrical resistivity measurements; blue dashed line is the pressure dependence of melting temperature. The results of quenching experiments are presented by circles (solid, half-filled and open symbols correspond to $B_6Si$, $B_6Si + B_{12}Si + Si$, and $B_nSi + Si$ ($12 \leq n \leq 50$) in the recovered samples, respectively). Red dashed line is the low-temperature boundary of $B_6Si$ melt disproportionation.